# THE MANGANITES AS METALLIC SEMICONDUCTORS

B.V. Karpenko[a] and A.V. Kuznetzov[b]

a) Karpenko Boris Victorovich, Institute of Physics of Metals, URO RAN, Ekaterinburg 620990, Russia, phone: (343) 378-35-64, fax: (343) 374-52-44, e-mail: boris.karpenko@mail.ru
b) Kuznetzov Alexander Vasilyevich, Ural State University, Ekaterinburg 620083, Russia, phone: (343) 218-05-77, e-mail: al.vas.kuz@gmail.com

## Abstract

The two-band model possessing metal and semiconductor properties is offered. This model may explain the appearance of temperature maximum of the electrical resistance at the proper relationships between the parameters of the theory. The calculations are carried in the tight-binding approximation. The applicability of the model to the manganites is discussed.

It is known that ferromagnetic conducting manganites have electrical resistivity maximum at approximately Curie temperature. The low-temperature region they call the metallic part (the resistance rises with the temperature growth) and high- temperature region the semiconductor or dielectric part (the resistance falls with the temperature growth). Often this phenomenon is named metal-insulator transition. (The experimental data, discussions and corresponding references may be found in the papers [1-3].)

Below the simple two-band model of conductor will be considered. In the frame of this model the temperature maximum of electrical resistance can be obtained. This model can not give the complete picture the resistance mechanism in the manganites but can give certain contribution in this phenomenon.

Then let we have the system consisting of two electronic energy bands, lower and upper. In general these bands could be separated by the arbitrary energy gap. This gap may have positive and negative value (the case of overlapping bands). Then let the lower band has the arbitrary degree of electron filling at absolute zero temperature and the upper band is completely empty in absence of overlapping. It is seen that this system may correspond to simple metal, intrinsic semiconductor or semimetal in dependence of the gap value and filling degree.

Let us consider that the Fermi distribution is valid for both bands:

$$f_1 = \frac{\lambda_1}{1 + \exp\dfrac{E_1 - \mu}{kT}} \quad , \quad f_2 = \frac{\lambda_2}{1 + \exp\dfrac{E_2 - \mu}{kT}} \quad . \tag{1}$$

In Eq. (1) the indexes 1 relate to first (lower) band and indexes 2 relate to second (upper) band. We used the designations: $E$ is the band energy, $\mu$ is the chemical potential, $\lambda$ is spin degeneration, $k$ is Boltzmann constant, $T$ is temperature. Chemical potential $\mu$ is determined in usual manner:

$$\sum_{states} f_1 + \sum_{states} f_2 = N_e, \tag{2}$$

where $N_e$ is complete number of electrons.

Below we will make the following simplified assumptions and concretizations.

1. We assume that for both bands the tight-binding approximation for simple cubic lattice is valid.

2. The bands widths are equal one another.

3. Spin degeneration is removed.

Then we have in the nearest neighbor approximation and in zero electric field

$$E_1 = \Delta \varepsilon, \tag{3}$$

$$\varepsilon = \frac{1}{6}(3 - \cos x - \cos y - \cos z), \ -\pi \leq x, y, z \leq \pi, \tag{4}$$

$$E_2 = E_1 + \Delta + \delta, \tag{5}$$

$$\lambda_1 = \lambda_2 = 1. \tag{6}$$

(Here $\Delta$ is bandwidth, $\delta$ is energy gap; $x, y, z$ are the components of quasi-impulse.)

For the sake of simplicity let us introduce the dimensionless magnitudes::

$$t = \frac{kT}{\Delta}, \ w = \frac{\mu}{\Delta}, \ \gamma = \frac{\delta}{\Delta} . \tag{7}$$

In such case the Eq. (2) for chemical potential becomes

$$\frac{1}{(2\pi)^3} \int_{-\pi}^{\pi} \int_{-\pi}^{\pi} \int_{-\pi}^{\pi} \left( \frac{1}{1 + \exp\frac{1}{t}(\varepsilon - w)} + \frac{1}{1 + \exp\frac{1}{t}(\varepsilon + 1 + \gamma - w)} \right) dx\,dy\,dz = n, \tag{8}$$

Where $n$ is the concentration

$$n = \frac{N_e}{N} \tag{9}$$

and $N$ is the lattice sites number.

Below we will consider the temperature dependence of the chemical potential for special cases:

$$n = 0.7 \ \text{и} \ \gamma = 1; 0.1; 0.01; 0; -0.01; -0.1; -1. \tag{10}$$

The graphical solutions of Eq. (8) are represented in Fig. 1 for these parameters.

It is seen that concentration in lower band $n_1$ falls and concentration in upper band $n_2$ rises with the temperature for different $\gamma > -1$. For the case $\gamma = -1$ we have naturally $n_1 = n_2 = n/2 = 0.35$.

Further we have to calculate the electrical resistivity. The simplest time relaxation approximation will be used. Let $\tau_1$ and $\tau_2$ are the relaxation times of lower and upper band correspondingly. The standard solution of Boltzmann equation gives the following expression for the conductivity $\sigma$ :

$$\sigma = \frac{e^2 \Delta}{(2\pi)^3 \hbar^2 a} (\tau_1 K_1 + \tau_2 K_2).\tag{11}$$

In Eq. (11) we used the notations:

$e$ - electron charge, $\hbar$ - Planck constant, $a$ - lattice parameter;

$$K_1 = \frac{1}{t} \int_{-\pi}^{\pi} \int_{-\pi}^{\pi} \int_{-\pi}^{\pi} \frac{(\sin^2 x)\exp\frac{1}{t}(\varepsilon - w)}{[1+\exp\frac{1}{t}(\varepsilon - w)]^2} dxdydz,\tag{12}$$

$$K_2 = \frac{1}{t} \int_{-\pi}^{\pi} \int_{-\pi}^{\pi} \int_{-\pi}^{\pi} \frac{(\sin^2 x)\exp\frac{1}{t}(\varepsilon + 1 + \gamma - w)}{[1+\exp\frac{1}{t}(\varepsilon + 1 + \gamma - w)]^2} dxdydz,\tag{13}$$

For the numerical example we take

$$\tau_2 = 10\tau_1 \text{ и } \gamma = 1.\tag{14}$$

Using the values $w$ from Eq. (8) we obtain the graph of reduced resistance $r(t)$ presented in Fig. 3. (The temperature independence approximation of time relaxation was used.)

$$r(t) = \left(K_1 + 10K_2\right)^{-1}.\tag{15}$$

The physical meaning of the maximum on the curve $r(t)$ is quite understandable: at low temperatures due to presence of empty states the metallic conductivity character is realized ($r(t)$ rises with temperature). More over upper band practically does not affect on conductivity. As further temperature growth the upper band electrons become give their contribution, resistivity curve passes through the maximum and then it falls as in semiconductor. At high temperatures resistivity rises again owing to natural reasons. More over no metal-insulator transition arises. Simply such resistance behavior is natural. Therefore we call this system metallic semiconductor.

Let us discuss the obtained results from the point of view their applicability to the special case of the manganites. In the first our two-band model corresponds to presence of two bands $e_{g1}$ and $e_{g2}$ originated from splitting of $e_g$-states of the manganese ions. In the second above used the condition $\lambda_1 = \lambda_2 = 1$ means the spin polarization of the conductivity electrons which exists in the manganites due to double exchange mechanism: the $e_g$ electron spins are parallel to the large spins $S = 3/2$ of the localized manganese $t_{2g}$-electrons. We neglected the anti-Hund states of $e_g$-electrons at all. In the third above used concentration value $n = 0.7$ corresponds to the doping degree $x = 0.3$ ($n = 1 - x$) in the compounds $RE_{1-x}M_xMnO_3$ ($RE =$ rare earth, $M = Ca, Sr, Ba$). At this (or near it) concentration the ferromagnetic and metallic-semiconductor properties become highly apparent. At these points the direct correspondence of our model to the real manganites is finish. Choose at the dispersion law (Eqs. (3), (4) and (5)) is arbitrary and valid only in the tight binding approximation for nearest neighbors. The more realistic dispersion law is needed. In used approximation of the relaxation time all real scattering mechanisms (scattering on phonons, localized spins and other) are hidden in the parameter $\tau_1$ and $\tau_2$, therefore the real temperature resistivity dependence will be other than in Eq.(15) and on Fig. 3. The absence of any connection between the Curie temperature and resistivity maximum is the shortage of the present model. In the proposed model there are enter such parameters as lower bandwidth, upper bandwidth, magnitude of energy gap, two relaxation times and therefore it possible to obtain practically any shape for the resistivity varying these parameters. In the present work we did not used the numerical values for the bandwidths and energy gap because their inaccessibility.

Let us note that we used the spectrum with the energy minimum in the center of Brillouin zone for the both bands. However the same results are valid for the cases when the maximum is in center of zone for the bands and also minimum or maximum for different bands (four variants in all).

Let us say that in spite of all shortages of the model the main features of the present metallic-semiconductor model have to be used in more perfect theories of manganites.

### Acknowledgements

The author is indebted OFN RAN N 09-T2-1013, URO-SO RAN N 09-S2-1016.

**Figure captions**

Fig.1.The temperature dependence of chemical potential $w$ for parameter $\gamma = 1; 0.1; 0.01; 0; -0.01; -0.1; -1$ (from the top).

Fig.2.The temperature dependences of the concentration $n_1$ (the descending curves) and $n_2$ (the ascending curves) for different $\gamma$. The horizontal curves correspond to full concentration $n = 0.7$.

Fig.3.The temperature dependence of the reduced resistance $r$.

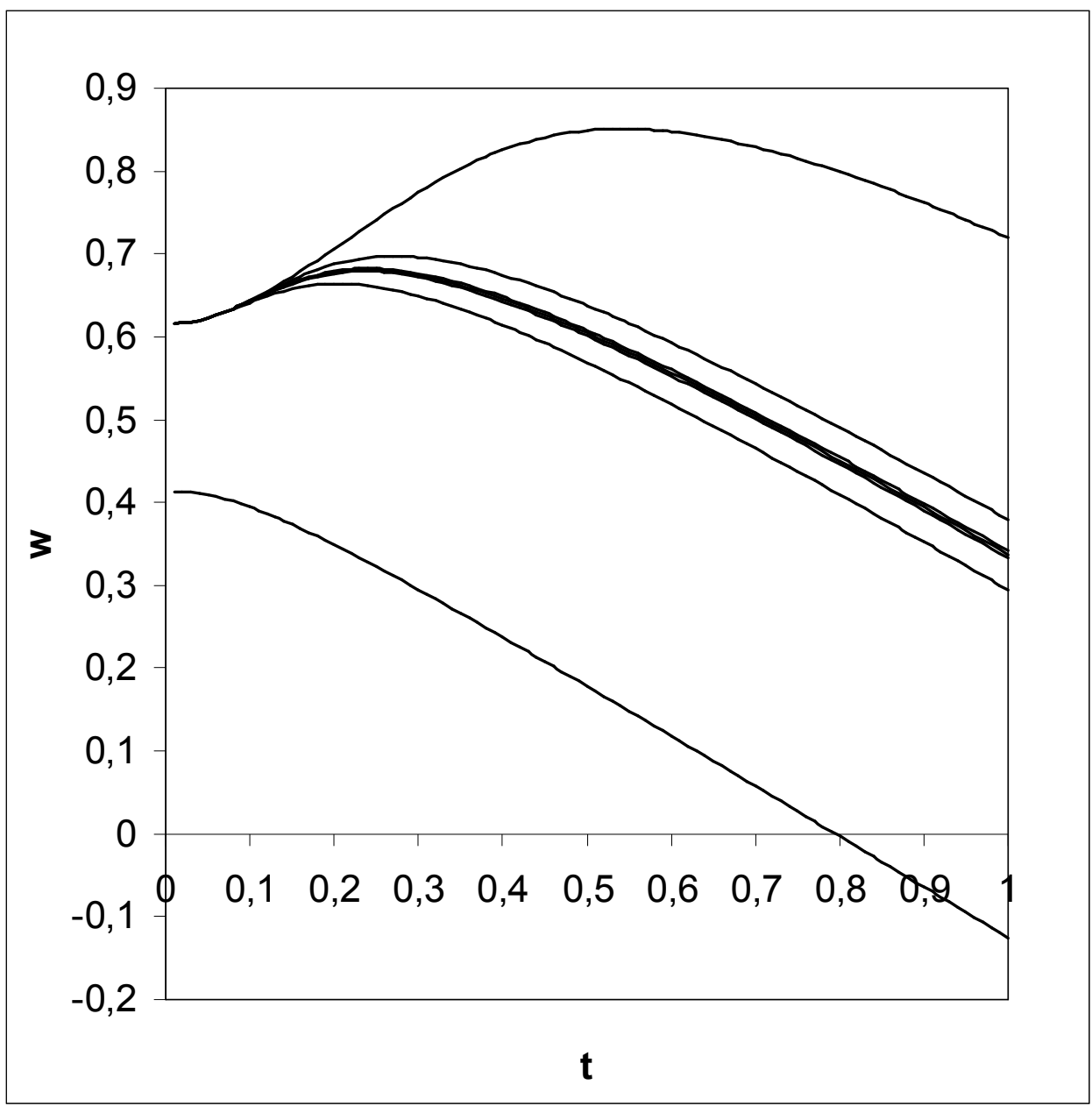

**Fig.1. B.V. Karpenko and A.V. Kuznetsov, THE MANGANITES AS METALLIC SEMICONDUCTOR**

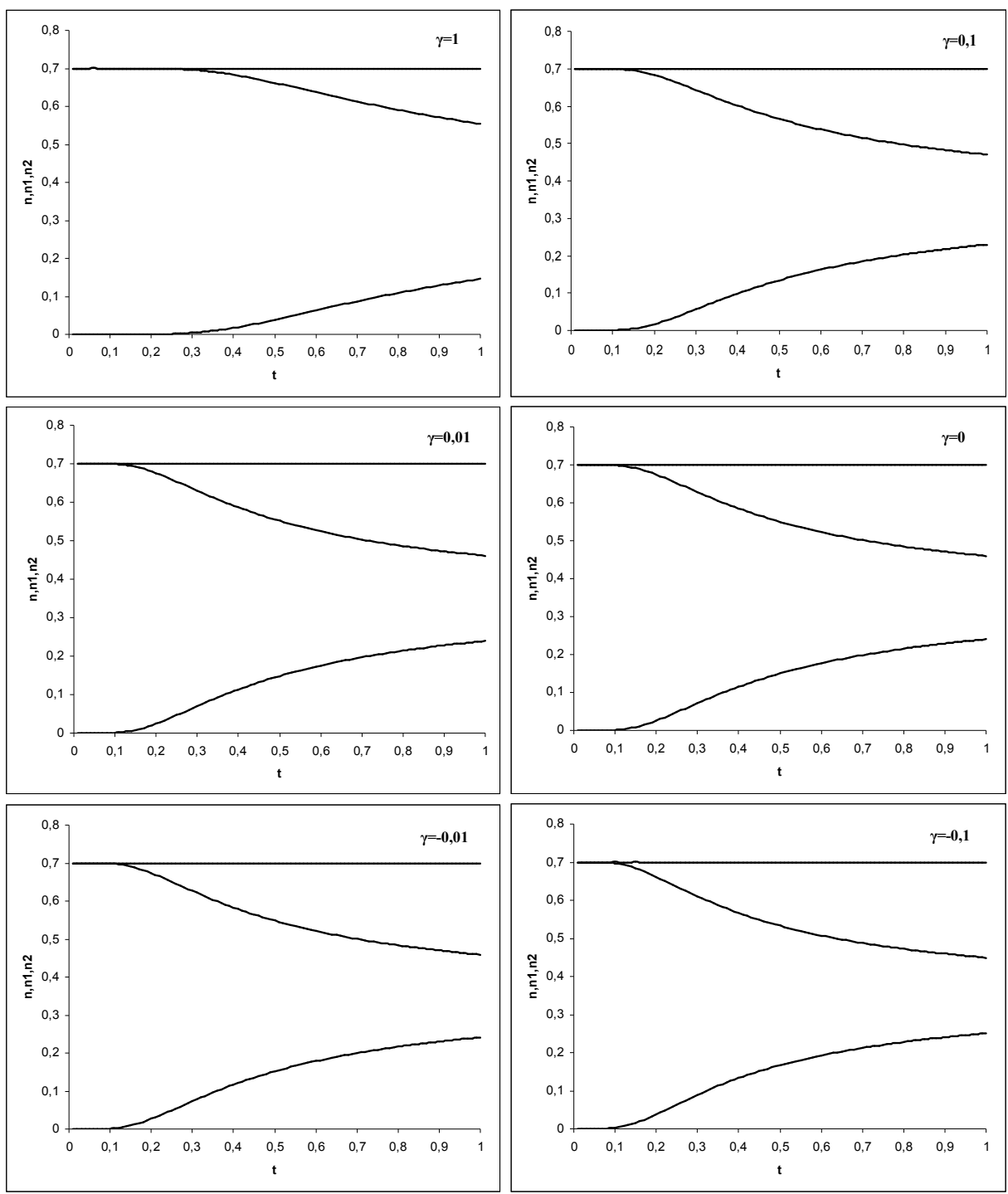

**Fig .2. B.V. Karpenko and A.V. Kuznetsov, THE MANGANITES AS METALLIC SEMICONDUCTORS**

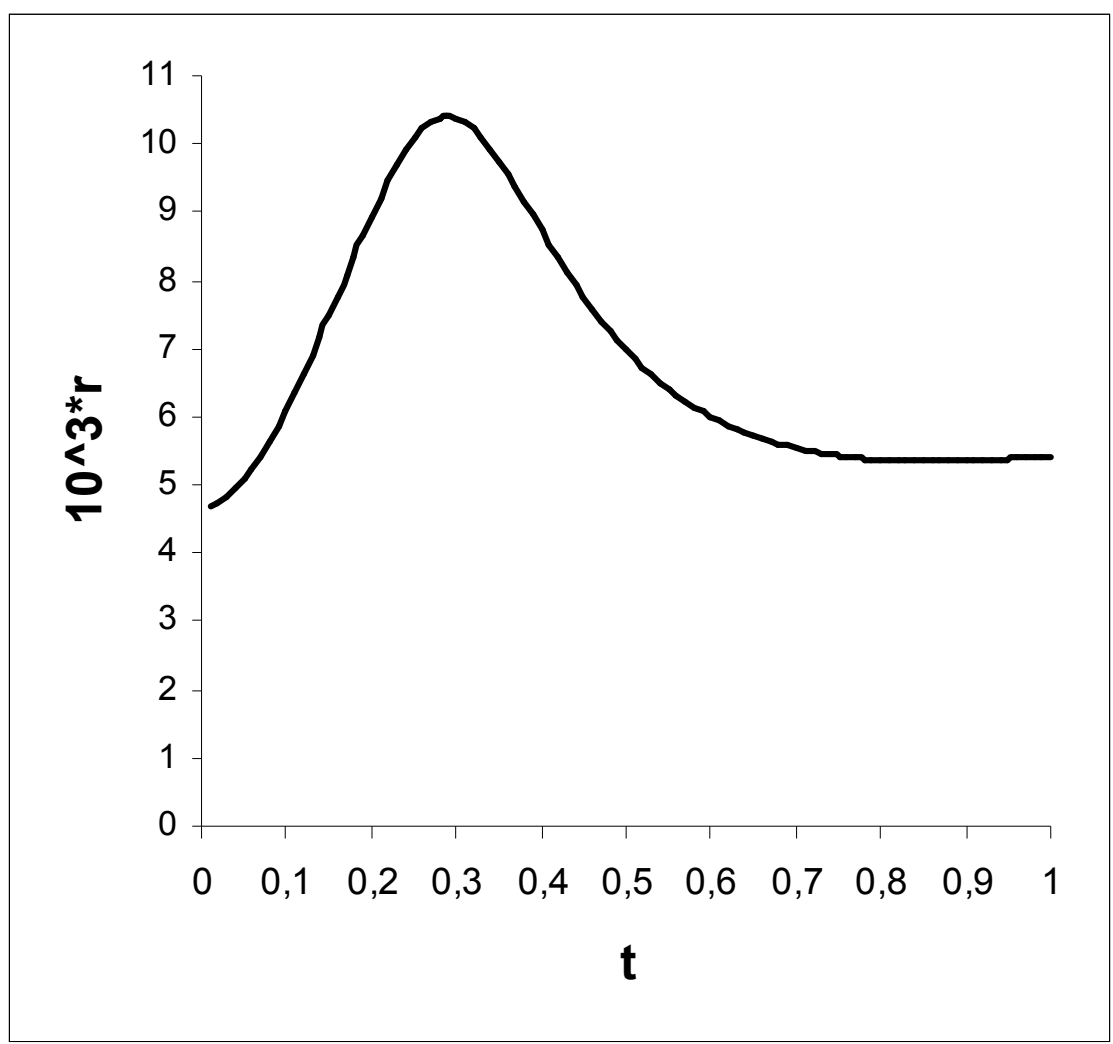

**Fig.3. B.V. Karpenko and A.V. Kuznetsov, THE MANGANITES AS METALLIC SEMICONDUCTORS**